\documentclass[conference]{IEEEtran}
\IEEEoverridecommandlockouts
\usepackage[utf8]{inputenc}
\usepackage[T1]{fontenc}
\usepackage{fix-cm}
\usepackage{import}
\usepackage{ifpdf}
\usepackage{lscape}
\usepackage[pdftex]{graphicx}
\usepackage{epstopdf}
\usepackage[table]{xcolor}
\usepackage{soul}
\usepackage[cmex10]{amsmath}
\usepackage[binary-units, per-mode=symbol]{siunitx}
\usepackage{array}
\usepackage{mdwmath}
\usepackage{diagbox}

\usepackage[tight,footnotesize]{subfigure}
\usepackage{listings}
\usepackage{caption}
\usepackage{framed}
\usepackage[many]{tcolorbox}
\usepackage{float}
\usepackage{flushend}
\usepackage[nolist]{acronym}
\usepackage{cite}
\usepackage[bottom]{footmisc}
\usepackage{fixfoot}
\usepackage[disable]{todonotes}
\usepackage{marginnote}
\usepackage{url}
\usepackage{hyperref}
\usepackage{lipsum}

\newfloat{lstfloat}{htbp}{lop}
\floatname{lstfloat}{Listing}
\restylefloat{lstfloat}

\definecolor{dkgreen}{rgb}{0, 0.6, 0}
\definecolor{specgray}{rgb}{0.92, 0.92, 0.92}
\definecolor{mauve}{rgb}{0.58, 0, 0.82}

\let\origthelstnumber\thelstnumber
\makeatletter
\newcommand*{\Suppressnumber}{%
	\lst@AddToHook{OnNewLine}{%
		\let\thelstnumber\relax%
		\advance\c@lstnumber-\@ne\relax%
	}%
}

\newcommand*{\Reactivatenumber}{%
	\lst@AddToHook{OnNewLine}{%
		\let\thelstnumber\origthelstnumber%
		\advance\c@lstnumber\@ne\relax%
	}%
}

\presetkeys{todonotes}{size=\tiny}{}

\newcommand{\margintodo}[2][]{%
	\let\savemarginpar\marginpar%
	\let\marginpar\marginnote%
	\todo[#1]{#2}%
	\let\marginpar\savemarginpar%
}

\newenvironment{nestingsection}[1]{%
	\section{#1}%
	\let\section\subsection%
	\let\subsection\subsubsection%
	\let\subsubsection\paragraph%
	\let\paragraph\subparagraph%
	\let\subparagraph\relax%
}

\begin{document}

\title{A Short Scalability Study on the SeQUeNCe Parallel Quantum Network Simulator
}

\author{
	\IEEEauthorblockN{Aaron Welch}
		\IEEEauthorblockA{Computational Sciences \& \\Engineering Division}
		\IEEEauthorblockA{Oak Ridge National Laboratory, USA\\
		welchda@ornl.gov}
	\and
	\IEEEauthorblockN{Mariam Kiran}
		\IEEEauthorblockN{Computational Sciences \& \\Engineering Division}
		\IEEEauthorblockA{Oak Ridge National Laboratory, USA\\
		kiranm@ornl.gov}
}

\maketitle

\begin{abstract}
As quantum networking continues to grow in importance, its study is of interest to an ever wider community and at an increasing scale.
However, the development of its physical infrastructure remains burdensome, and services providing third-party access are not enough to meet demand.
A variety of simulation frameworks provide a method for testing aspects of such systems on commodity hardware, but are predominantly serial and thus unable to scale to larger networks and/or workloads.
One effort to address this was focused on parallelising the SeQUeNCe discrete event simulator, though it has yet to be proven to work well across system architectures or at larger scales.
Therein lies the contribution of this work --- to more deeply examine its scalability using \acs{ORNL}'s Frontier.
Our results provide new insight into its scalability behaviour, and we examine its strategy and how it may be able to be improved.

\end{abstract}


\begin{nestingsection}{Introduction}
\label{sec:intro}
Quantum networks are argued to be the next generation for communication networks that will deliver entanglement and connect distributed quantum physics devices such as quantum computers, sensors, and detectors \cite{chiribella2009theoretical, wei2022towards, kozlowski2019towards, Peters:2023mpv}.
Instead of classical ``bits'', qubits are transmitted that encode information using the photon spin representation.
Quantum networks are being designed to develop ultra-secure and highly accurate sensor networks for science and commercial applications \cite{doi:10.1021/acsphotonics.9b00250}.

There are numerous industry efforts developing new hardware, protocols, and tools to enable quantum information exchange \cite{muralidharan2016optimal}, using discrete and continuous variable demonstrations, each bringing their own capabilities \cite{Alshowkan:2021kpt}, fiber-optical transmission or free space point-to-point communication.
However, network loss plays a huge role in guaranteeing the validity of the quantum states, with current implementations limited to a \SI{300}{\kilo\meter} distance \cite{Rao:2023exv}.
Using quantum repeaters, one can extend these to longer distances such that quantum states can be refreshed or preserved \cite{azuma2023quantum, singh2021quantum, desef2021protecting}.

In order to design more efficient quantum networks and provide a cheap alternative for testing and analysis, quantum network simulations have been developed to help define use cases, and collect data to build reliable devices.
However, these have been largely sequential and thus limited to small networks and time scales.
SeQUeNCe~\cite{sequence} addresses this limitation by enabling parallel discrete event simulation~\cite{10.1145/76738.76741} that can scale across many processes or nodes \cite{10.1145/3634701}.
It achieves a high level of customisability with its modular design, which comprises six components --- the simulation kernel, the hardware, entanglement management, network management, and resource management modules, and the application module.

However, the scalability of parallel execution with SeQUeNCe has only been explored in a few studies~\cite{10.1145/3384441.3395988,10.1145/3634701}, leaving questions about its practical use for large-scale quantum network simulations.
This paper's key contribution is investigating this issue, with a primary focus on the scalability of the simulation kernel, which includes an event scheduler and a \ac{QSM}.
The kernel processes the simulation along a discrete timeline, with events generated by the other modules stored within a priority work queue, where the event with the earliest timestamp is repeatedly removed and executed until the simulation is rendered complete.
The \ac{QSM} manages all quantum states using a key-value store, which may only be manipulated through requests to its \acs{API}, ensuring simulation consistency.

This paper is organised as follows: Section~\ref{sec:background} highlights some other relevant simulation frameworks and provides an overview of SeQUeNCe itself and its parallelisation effort.
Section~\ref{sec:results} presents the results of our scalability studies and includes a brief analysis of the identified bottlenecks.
Finally, Section~\ref{sec:conclusion} contains our concluding remarks and suggestions for what may be required to overcome the scalability issues we encountered.

\end{nestingsection}

\begin{nestingsection}{Background and Related Work}
\label{sec:background}
A wide variety of other simulators have emerged in recent years, covering a wide range of use cases.
Several efforts such as QuNetSim~\cite{diadamo2021qunetsim}, ComNetsEmu~\cite{fitzek2020computing}, or Cisco's QNetLab~\cite{CISCO} are developing kits to interface with other simulations, and investigate new protocols.
Some other discrete event simulators include OMNET++~\cite{varga2010omnet++}, NetSim~\cite{veith1999netsim}, SimQN~\cite{chen2023simqn}, Netsquid~\cite{coopmans2021netsquid}, and NS-3~\cite{carneiro2010ns}, a network simulator that is frequently used in research for communication networks.
Alternatively, some opt to employ equation-based modelling, such as \ac{SQUANCH}~\cite{bartlett2018distributed}.

\ac{SQUANCH} is notable for being another simulator with support for parallel execution.
However, since it is based on approximate models, it has the potential to suffer from inaccuracies compared to what could be observed with physical counterparts to that being simulated.
QuISP~\cite{satoh2022quisp} is based on OMNET++ and is also relevant for its ability to incorporate noisy conditions to experiments in order to provide an accurate view of how systems may behave in more realistic settings.
While this can be useful, several repeated runs may be necessary as a consequence so as to increase confidence in the results.

\begin{nestingsection}{SeQUeNCe}
\label{sec:sequence}

Designed for use in the study of quantum networks, SeQUeNCe~\cite{sequence} has been released as an open source full-stack discrete event quantum network simulator written in Python that aims to have very fine precision.
Its structure is similar to the one used by NS-3.
The simulator explores hybrid classical and quantum networks, and also offers the ability to use classical control software to enable rapid experimental progress.

The drawback of the simulator is that it assumes no loss and perfect reliability for the classical channels.
This implies that its behaviour becomes uncertain if the network becomes saturated or if any network failures occur.
Additionally, it assumes that the host node is the initiator of any request sent in the application module.
Essentially, this means that having responses sent back to the initiator of a request may need to have some modifications performed in the application module of the simulator.

\end{nestingsection}
\begin{nestingsection}{Parallelising SeQUeNCe}
\label{sec:sequence-parallelisation}

Preliminary work exploring the parallelisation potential exploited within SeQUeNCe \cite{10.1145/3384441.3395988} uncovered five key observations based on an analysis of sequential simulation for \ac{QKD}:

\begin{itemize}
\item Events on quantum channels are dominant based on both quantity and simulation execution time.
\item The execution time for these events is highly consistent.
\item Events between pairs of \ac{QKD} terminals are evenly distributed over the simulated timeline.
\item The latency of quantum channel transmission is dominated by its propagation delay, which is lower than that of an equivalent classical channel.
\item Different \ac{QKD} sessions are largely independent.
\end{itemize}

Based on these observations, a parallel version of SeQUeNCe was developed to efficiently scale these types of simulations \cite{10.1145/3634701}.
It uses the simulated routers as the unit of parallelism, splitting the timeline into \(n\) parallel timelines.
Additionally, the \ac{QSM} is broken up into a hierarchy of \(n\) local \acp{QSM} that each manage quantum states which remain confined exclusively to the local process, and one global \ac{QSM} that is responsible for cohesively managing entangled states.
Communication between simulation processes occurs through \acs{MPI}, whereas requests for the global \ac{QSM} are handled through \acs{TCP}/\acs{IP} sockets.

To ensure that the order of events remains correct and consistent with serial execution, the parallel timelines must be subdivided into synchronisation epochs based on lookahead values that are short enough to guarantee event causality.
These lookahead calculations and resulting epochs continue to be computed until the simulation is complete across all timelines.
Within each epoch, processes begin by synchronising their timelines via \acs{MPI}, exchanging events, and computing the lookahead value that demarcates the end of the next epoch.
They then proceed to process all local events with timestamps within the epoch, batching requests to the global \ac{QSM}, which are then processed in bulk before moving on to the next epoch.

A critical factor in achieving scalable performance is maintaining a reasonable workload balance across all parallel timelines, as any imbalance will inherently lead to the straggler problem, where other processes must wait for the straggler to complete its computation before the synchronisation phase can begin.
As a result, extensive investigation has been conducted into developing and analysing different network partitioning schemes (i.e., determining which simulated routers are assigned to specific processes or nodes).
In particular, simulated annealing was used with energy functions optimising for cross-process flows, cross-process quantum channels, or the quantity of memories in each router.

However, in addition to being fairly expensive on its own, this kind of approach can reduce the generality of the simulator by requiring potentially unreasonably intimate knowledge of the simulation workload before the simulation even begins.
Furthermore, the scalability for even the best distributions will eventually still be bounded by the workload differential, so while our work still reproduced the same simulated annealing process, it was limited to focusing only on use of the simulated network topology for determining its partitioning layout.

The \ac{SST}~\cite{10.1145/1964218.1964225} is a discrete event simulator with a networking component for \acs{HPC} systems that received its own parallelisation effort.
While not focused on quantum technology nor sharing constraints uniquely presented by it, it may be able to provide some wisdom for improving upon the effort in SeQUeNCe.

\end{nestingsection}

\end{nestingsection}

\begin{nestingsection}{Experimental Studies}
\label{sec:results}
For our scalability studies, we reproduced two \ac{QKD} tests from \cite{10.1145/3634701} on \acs{ORNL}'s Frontier supercomputer, using the same input data and scripts.
As such, we simulated 1024 routers across both linear and \ac{AS} network topologies, distributed across up to 512 processes on 8 compute nodes, plus an additional node used exclusively for the global \ac{QSM}.
After the initial reproduction verifying prior results, this would have been scaled up further had the simulation not already hit limits in its scalability design.

\begin{nestingsection}{Testing Environment}
\label{sec:testing-environment}

The Frontier system has 74 \acs{HPE} Cray EX Olympus racks, each having 128 \acs{AMD} compute nodes, totaling 9,472 compute nodes.
Each compute node has a single 64-core \acs{AMD} Epyc 7A53 \SI{2}{\giga\hertz} \acs{CPU} with \SI{512}{\giga\byte} of memory and four \acs{HPE} Slingshot 11 \SI{200}{\giga\bit\per\second} Casinni \acsp{NIC} connected to four \acs{AMD} MI250X \acsp{GPU}.
We used Python 3.10 for the run-time environment.

Our testing utilised the Python version of the global \ac{QSM} instead of the also included C++ server.
This choice was made to evaluate the impact of each component relative to the number of processes/nodes, which might be less apparent with the C++ server.
That is, while the Python version is single-threaded, the C++ version is multi-threaded with a new thread spawned for each process connected to it.
Using the Python server allows us to more easily examine scaling patterns as a factor of the number of connected processes while still being straightforward to intelligbly improve upon as limits may start being hit.
However, as our results showed, performance was being held back by a far bigger problem before communication with the global \ac{QSM} could become an issue.

\end{nestingsection}

\begin{nestingsection}{Linear Topology}
\label{sec:linear-topology}

We tested a linear topology (as shown in Figure~\ref{fig:linear-topology}), in which all routers are connected in a single line, and the payload is to send requests from one end of the network to the other.
Here, we ran up to 128 processes using strong scaling, though due to the higher core count on Frontier, this resulted in spreading to multiple nodes after 64 processes rather than 32 as in \cite{10.1145/3634701}.

\begin{nestingsection}{Observations}
\label{sec:linear-observations}

We observed results similar to those in \cite{10.1145/3634701}, as seen in Figure~\ref{fig:linear-strong}.
Here, we break down the average time spent into three categories -- computing local events, communicating to the global \ac{QSM} (labelled ``socket''), and the subsequent synchronisation of timelines and cross-process events via \acs{MPI}.
This provides a degree of validation, though also meant that it showed the same apparent signs of plateauing that contributed to inspiring our study, with scaling appearing to cease by the time the first node is fully subscribed.
However, we note that such a simulated topology and workload is not very practical, so rather than extending, studying, and improving upon these results, we instead move next to a more realistic simulation in the form of an \ac{AS} network.

\begin{figure}[htbp]
\begin{minipage}{0.24\textwidth}
\begin{center}
\includegraphics[width=0.24\textwidth]{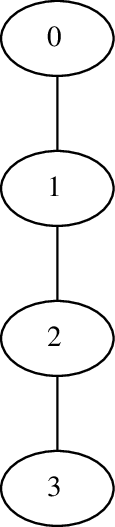}
\end{center}
\caption{Linear Topology}
\label{fig:linear-topology}
\end{minipage}
\begin{minipage}{0.24\textwidth}
\begin{center}
\includegraphics[width=\textwidth]{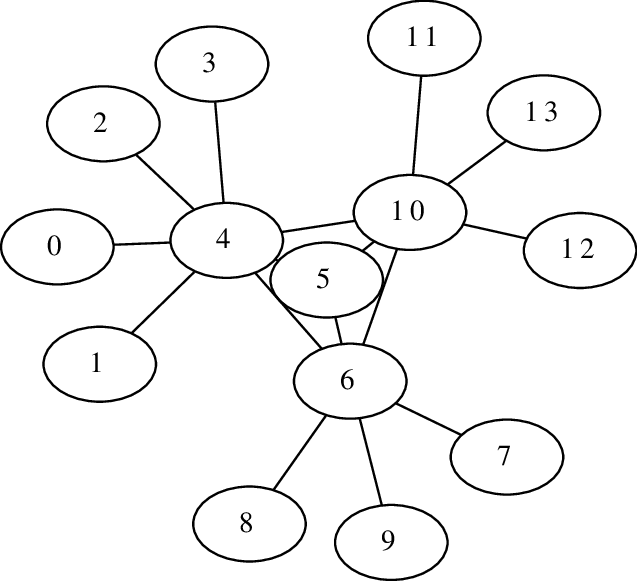}
\end{center}
\caption{Autonomous System Topology}
\label{fig:as-topology}
\end{minipage}
\end{figure}

\begin{figure}[htbp]
\begin{center}
\resizebox{0.5\textwidth}{!}{\subimport*{images/}{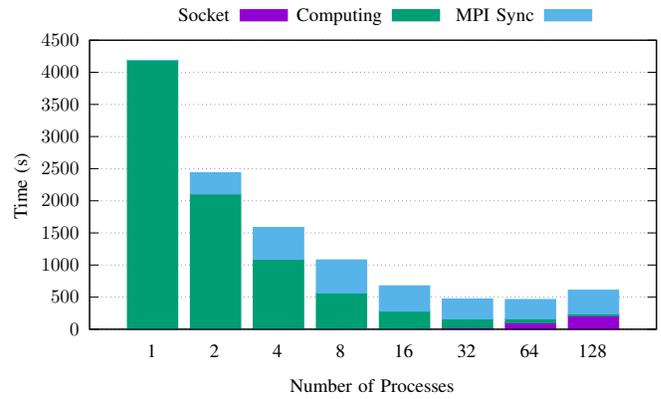}}
\end{center}
\caption{Linear Topology Performance}
\label{fig:linear-strong}
\end{figure}

\end{nestingsection}
\end{nestingsection}

\begin{nestingsection}{Autonomous System Topology}
\label{sec:as-topology}

An \ac{AS} network topology (as shown in Figure~\ref{fig:as-topology}) mirrors a common structure seen in global networks.
For this test, a more varied workload is spread across the network, which is partitioned based upon the cross-process quantum channels.
As the script used for generating the input data could not be made to output a single router per process, the process count used ranged from 1--512 (again using strong scaling), spread across up to 8 nodes.
The results of this can be seen in Figure~\ref{fig:as-net-qc}.
By extending the process count further, we can better confirm a disturbing trend --- that the performance continues to steadily degrade and become worse, even prior to fully utilising just a single node.
In fact, maximum performance was obtained at a mere 16 processes, after which increasing synchronisation costs began to outweigh the decreased computational burden.

\begin{figure}[htbp]
\begin{center}
\resizebox{0.5\textwidth}{!}{\subimport*{images/}{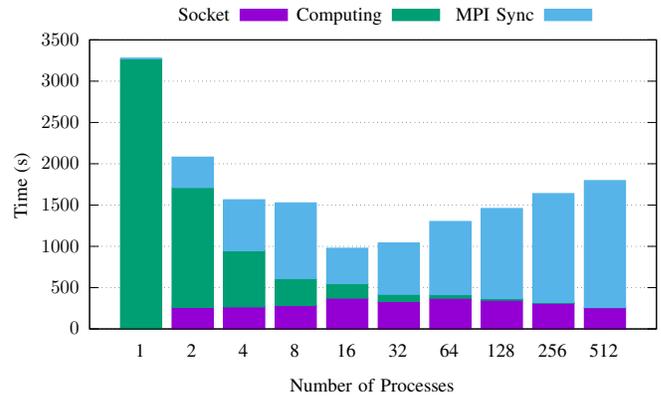}}
\end{center}
\caption{Autonomous System Performance}
\label{fig:as-net-qc}
\end{figure}

Upon investigating these synchronisation costs further, we note that the original timing methodology includes everything from the end of local event processing to the end of the next synchronisation phase, which could obscure how much time is lost simply due to waiting for stragglers on an unbalanced workload.
We therefore add a barrier prior to the actual communication and use it to split the time between what is spent waiting for other processes to be ready and actually engaging in the necessary communication, the results of which can be seen in Figure~\ref{fig:as-net-qc-breakdown}.

\begin{figure}[htbp]
\begin{center}
\resizebox{0.5\textwidth}{!}{\subimport*{images/}{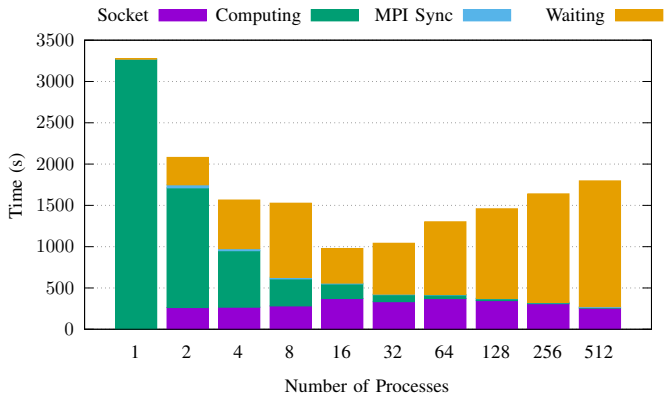}}
\end{center}
\caption{Autonomous System Performance Breakdown}
\label{fig:as-net-qc-breakdown}
\end{figure}

\begin{nestingsection}{Observations}
\label{sec:as-observations}

We can now determine that the network communication was never actually a significant concern as results previously suggested in \cite{10.1145/3634701}.
Instead, our results show that the apparent decrease in average computing time actually comes from the addition of relatively idle workers without any change to the overall workload or improvements to the straggling workers that remain with a dramatically larger share of it.
The simulator leaves much of the responsibility for a balanced workload on the user input data, since it is limited in its ability to mitigate such imbalance with partitioning alone.
Since we argue that time spent waiting on other processes should be included in determining overall compute time, Figure~\ref{fig:as-net-qc-redefined} adjusts the visualisation to account for this, which more accurately portrays the limitations of its scalability..

\begin{figure}[htbp]
\begin{center}
\resizebox{0.5\textwidth}{!}{\subimport*{images/}{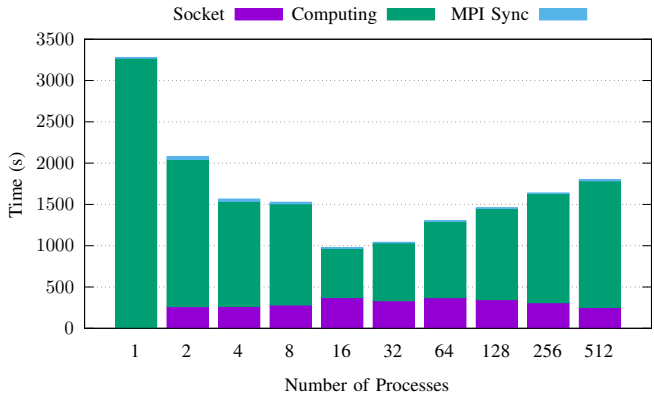}}
\end{center}
\caption{Autonomous System Performance Redefined}
\label{fig:as-net-qc-redefined}
\end{figure}

With the current simulator design, one can do little to further optimise this, since the unit of parallelism is the simulated router and that is precisely where the problem is.
We can see this in greater detail if we track the computing time within each synchronisation epoch for each simulation process, as can be seen for our 8 process run in Figure~\ref{fig:as-net-qc-computing-8}.

\begin{figure}[htbp]
\begin{center}
\resizebox{0.5\textwidth}{!}{\subimport*{images/}{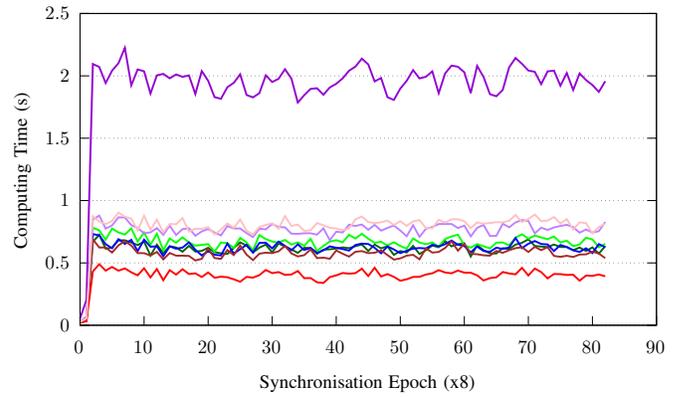}}
\end{center}
\caption{Autonomous System Computing Time Per Process}
\label{fig:as-net-qc-computing-8}
\end{figure}

Here, we collapsed every 8 epochs into a single data point and show computing time only, with each process represented by its own line.
The imbalance problem can be clearly seen with one process dominating the rest by a wide margin, and the overall situation can not be guaranteed to improve with better partitioning across larger process counts.
This suggests that an entirely new parallelisation strategy may be required in order to achieve and maintain better scaling for general simulation requirements.

\end{nestingsection}
\end{nestingsection}

\end{nestingsection}

\begin{nestingsection}{Conclusion}
\label{sec:conclusion}
While some degree of speedup potential was observed, the overall scalability results and the generality to which they could be applied did not meet expectations.
Scaling beyond a single node may not be likely to yield worthwhile improvements for sufficiently unbalanced workloads, which we expect could be typical of many common and realistic settings.
This would also obviate the need for having a separate process to act as a global \ac{QSM} if enacting a purely multithreaded approach over the current multi-process design, eliminating another potentially significant source of performance bottlenecking.
Such a change may also present more effective opportunities for dynamic work sharing/stealing.
Even for balanced workloads, implementing a hybrid parallelism approach with all work on a single node being done with multiple threads within a single process could dramatically reduce synchronisation costs and load on the global \ac{QSM}.

Furthermore, since the unit of parallelism is the router, scaling up to increasingly larger process/node counts requires simulating equally larger numbers of routers, for which the threshold beyond which an equivalent configuration in a real world scenario no longer makes sense could be quickly reached.
We see a need to redesign the parallelisation strategy itself, which could provide a better opportunity to efficiently scale not just the topology, but also the workload applied to it as well.

\end{nestingsection}


\begin{acronym}
\acro{AGAS}{active global address space}
\acro{AGP}{atomic, get, and put}
\acro{AI}{artificial intelligence}
\acro{AIMD}{ab initio molecular dynamics}
\acro{ALPS}{Application Level Placement Scheduler}
\acro{AMD}{Advanced Micro Devices}
\acro{AMO}{atomic memory operation}
\acrodefplural{AMO}[AMOs]{atomic memory operations}
\acro{AMR}{adaptive mesh refinement}
\acro{APGAS}{Asynchronous Partitioned Global Address Space}
\acro{API}{Application Programming Interface}
\acrodefplural{API}[APIs]{Application Programming Interfaces}
\acro{ARMCI}{Aggregate Remote Memory Copy Interface}
\acro{AS}{autonomous system}
\acro{AST}{abstract syntax tree}
\acro{BCL}{Berkeley Container Library}
\acro{BO}{Born-Oppenheimer}
\acro{CAF}{Coarray Fortran}
\acro{CAL}{Chapel Aggregation Library}
\acro{CAS}{compare and swap}
\acro{CCX}{Core CompleX}
\acrodefplural{CCX}[CCXs]{Core CompleXes}
\acro{CI}{continuous integration}
\acro{CPU}{central processing unit}
\acrodefplural{CPU}[CPUs]{central processing units}
\acrodefplural{CRDP}[CRDP]{Computational Research and Development Programs}
\acro{CSR}{compressed sparse row}
\acro{CSV}{comma-separated values}
\acro{DARPA}{Defense Advanced Research Projects Agency}
\acro{DBCSR}{distributed block compressed sparse row}
\acro{DCT}{Dynamically Connected Transport}
\acro{DDR}{double data rate}
\acro{DFT}{density-functional theory}
\acro{DoD}{Department of Defense}
\acro{DoE}{Department of Energy}
\acro{DPU}{data processing unit}
\acrodefplural{DPU}[DPUs]{data processing units}
\acro{DRAM}{dynamic random access memory}
\acro{DRASync}{Distributed Region-based Allocation and Synchronization}
\acro{ESSC}{Extreme Scale Systems Center}
\acro{FAM}{Fabric-Attached Memory}
\acro{FDO}{feedback directed optimisation}
\acrodefplural{FDO}[FDOs]{feedback directed optimisations}
\acro{FFTW}{fastest fourier transform in the west}
\acrodefplural{FPGA}[FPGAs]{field-programmable gate array}
\acro{GA}{Global Arrays}
\acro{GAiN}{Global Arrays in NumPy}
\acro{GAS}{global address space}
\acro{GASPI}{Global Address Space Programming Interface}
\acro{GCC}{GNU Compiler Collection}
\acro{GIL}{Global Interpreter Lock}
\acro{GPU}{graphics processing unit}
\acrodefplural{GPU}[GPUs]{graphics processing units}
\acro{GUI}{graphical user interface}
\acro{GUPS}{Giga UPdates per Second}
\acro{HBM}{high bandwidth memory}
\acro{HCA}{host channel adapter}
\acro{HCLib}{Habanero-C Library}
\acro{HDR}{high data rate}
\acro{HPC}{high performance computing}
\acro{HPCC}{High Performance Computing Challenge}
\acro{HPCS}{High Productivity Computing Systems}
\acro{HPE}{Hewlett Packard Enterprise}
\acro{HPF}{High Performance Fortran}
\acro{IB}{InfiniBand}
\acro{IBM}{International Business Machines}
\acro{IBV}{InfiniBand verbs}
\acro{IEEE}{Institute of Electrical and Electronics Engineers}
\acro{IP}{Internet Protocol}
\acro{IR}{intermediate representation}
\acro{I/O}{input/output}
\acro{JIT}{just in time}
\acro{LANL}{Los Alamos National Laboratory}
\acro{LLVM}{low level virtual machine}
\acro{LS-SCF}{linear-scaling self-consistent field}
\acro{MESI}{Modified, Exclusive, Shared, Invalid}
\acro{MPI}{Message Passing Interface}
\acro{MPT}{Message Passing Toolkit}
\acro{MXM}{MellanoX Messaging}
\acro{NCCL}{NVIDIA Collective Communication Library}
\acro{NIC}{network interface card}
\acro{NPB}{NAS Parallel Benchmarks}
\acro{NUMA}{non-uniform memory access}
\acro{NVM}{non-volatile memory}
\acro{OFA}{Open Fabrics Alliance}
\acro{OLCF}{Oak Ridge Leadership Computing Facility}
\acro{ORNL}{Oak Ridge National Laboratory}
\acro{ORTE}{Open Run-Time Environment}
\acro{OSU}{Ohio State University}
\acro{PAMI}{Parallel Active Messaging Interface}
\acro{PAPI}{Performance Application Programming Interface}
\acro{PCIe}{Peripheral Component Interconnect Express}
\acro{PCP}{Parallel C Preprocessor}
\acro{PE}{processing element}
\acrodefplural{PE}[PEs]{processing elements}
\acro{PGAS}{partitioned global address space}
\acro{PNNL}{Pacific Northwest National Laboratory}
\acro{POSH}{Python Object Sharing}
\acro{PPN}{processes per node}
\acro{QDNS}{Quantum Dynamic Network Simulator}
\acro{QDR}{quad data rate}
\acro{QKD}{quantum key distribution}
\acro{QSM}{quantum state manager}
\acrodefplural{QSM}[QSMs]{quantum state managers}
\acro{QoS}{quality of service}
\acro{RDMA}{remote direct memory access}
\acro{RMA}{remote memory access}
\acro{RTE}{runtime environment}
\acro{SCF}{self-consistent field}
\acro{SCoP}{static control part}
\acrodefplural{SCoP}[SCoPs]{static control parts}
\acro{SGI}{Silicon Graphics, Inc.}
\acro{SHARP}{Scalable Hierarchical Aggregation and Reduction Protocol}
\acro{SHOC}{Scalable Heterogeneous Computing}
\acro{SIMD}{single instruction, multiple data}
\acro{SLURM}{Simple Linux Utility for Resource Management}
\acro{SmartNIC}{smart network interface card}
\acrodefplural{SmartNIC}[SmartNICs]{smart network interface cards}
\acro{SPMD}{single program, multiple data}
\acro{SRS}{Storage Retrieval System}
\acro{SST}{Structural Simulation Toolkit}
\acro{SQL}{Structured Query Language}
\acro{SQUANCH}{Simulator for QUAntum Networks and CHannels}
\acro{STCI}{Scalable Tools Communication Infrastructure}
\acro{SVE}{Scalable Vector Extension}
\acro{TAU}{Tuning and Analysis Utilities}
\acro{TCP}{Transmission Control Protocol}
\acro{UCB}{University of California Berkeley}
\acro{UCCS}{Universal Common Communication Substrate}
\acro{UCP}{UC-Protocols}
\acro{UCS}{UC-Services}
\acro{UCT}{UC-Transports}
\acro{UCX}{Unified Communication X}
\acro{UH}{University of Houston}
\acro{UPC}{Unified Parallel C}
\acro{UTK}{University of Tennessee Knoxville}
\acro{VPI}{Virtual Protocol Interconnect}
\acrodefplural{WAN}[WANs]{wide area networks}
\acro{XML}{Extensible Markup Language}
\acro{XSLT}{Extensible Stylesheet Language Transformations}
\end{acronym}

\bibliographystyle{IEEETran}
\bibliography{main}

\end{document}